\documentclass[3p]{elsarticle}

\usepackage[utf8]{inputenc}
\usepackage{bm}
\usepackage{algorithm}
\usepackage[noend]{algpseudocode}
\usepackage{booktabs}
\usepackage{graphicx} 

\usepackage{mathtools}

\usepackage[english]{babel}

\usepackage{amssymb,amsmath} 

\usepackage{mathrsfs}

\usepackage{bbm}

\newtheorem{theorem}{Theorem}
\newtheorem{proposition}[theorem]{Proposition}
\newtheorem{corollary}[theorem]{Corollary}
\newtheorem{lemma}[theorem]{Lemma}
\newtheorem{definition}[theorem]{Definition}
\newtheorem{conjecture}[theorem]{Conjecture}

\algnewcommand\algorithmicforeach{\textbf{for each}}
\algdef{S}[FOR]{ForEach}[1]{\algorithmicforeach\ #1\ \algorithmicdo}

\begin{document}

\title{Minimum Spanning Tree Cycle Intersection Problem}

 \author[1,2]{Manuel Dubinsky\corref{cor1}}
 \ead{mdubinsky@undav.edu.ar}
 
\author[3]{César Massri \fnref{fn1}}

 \author[4]{Gabriel Taubin \fnref{fn2}}
 
\cortext[cor1]{Corresponding author}
\fntext[fn1]{Massri was supported by Instituto de Investigaciones Matemáticas ``Luis A. Santal\'o'', UBA, CONICET, CABA, Argentina}
\fntext[fn2]{Taubin was partially supported by NSF grant number IIS-1717355.}

\address[1]{Ingeniería en Informática, Departamento de Tecnología y Administración, Universidad Nacional de Avellaneda, Argentina}
\address[2]{Departamento de Computación, Facultad de Ciencias Exactas y Naturales, Universidad de Buenos Aires, Argentina}
\address[3]{Departamento de Matemática, Universidad de CAECE, CABA, Argentina}
\address[4]{School of Engineering, Brown University, Providence, RI, USA}

\begin{abstract}
Consider a connected graph $G$ and let $T$ be a spanning tree of $G$. 
Every edge $e \in G-T$ induces a cycle in $T \cup \{e\}$. The 
intersection of two distinct such cycles is the set of edges of 
$T$ that belong to both cycles. We 
consider the problem of finding a spanning tree  
that has the least number of such non-empty intersections.  In this article we analyze the particular case of complete graphs, and formulate a conjecture for graphs that have a universal vertex. \end{abstract}


\maketitle

\section{Introduction}

In this article we present what we believe is a new problem in 
graph theory, namely the Minimum Spanning Tree Cycle Intersection 
(\emph{MSTCI}) problem which arose while investigating a (yet unpublished) method for \emph{mesh deformation} in the area of \emph{digital geometry processing}, see \cite{Botsch:2010}. 

\bigskip

The problem can be expressed as follows. Let $G$ be a graph 
and $T$ a spanning tree of $G$. Every edge $e \in G-T$ induces a 
cycle in $T \cup \{e\}$. The intersection of two distinct such cycles 
is the set of edges of $T$ that belong to both cycles. Consider the 
problem of finding a spanning tree that has the least number of such 
pairwise non-empty intersections. 

\bigskip

The remainder of this section is dedicated to express the problem in the context of the theory of \emph{cycle bases}, where it has a natural formulation, and to describe an application. Section 2 sets some notation and convenient definitions. In Section 3 the complete graph case is analyzed. Section 4 presents a variety of interesting properties, and a conjecture in the slightly general case of a graph (not necessarily complete) that admits a star spanning tree. Section 5 explores programmatically the space of spanning trees to provide evidence that the conjecture is well posed. Section 6 collects the conclusions of the  article.

\subsection{Cycle bases}

The study of cycles of graphs has attracted attention for many years. To mention just three well known results consider \emph{Veblen's theorem} \cite{Veblen:1912} that characterizes graphs whose edges can be written as a disjoint union of cycles, \emph{Maclane's planarity criterion} \cite{Maclane:1937} which states that planar graphs are the only to admit a 2-basis, or the \emph{polygon matroid} in Tutte's classical formulation of \emph{matroid theory} \cite{Tutte:1965}.

\bigskip

The set of cycles of a graph has a vector space structure over $\mathbb{Z}_2$, in the case of undirected graphs, and over $\mathbb{Q}$, in the case of directed graphs \cite{Kavitha:2009}. A basis of such a vector space is denoted \emph{cycle basis} and its dimension is the \emph{cyclomatic number} $\nu = |E| - |V| + |CC|$ where $E$, $V$ ad $CC$ are the set of edges, vertices and connected components of the graph, resp. Given a cycle basis $B$ we can define its \emph{cycle matrix} $\Gamma \in K^{|E| \times \nu}$ where $K$ is the scalar field (i.e.: $\mathbb{Z}_2$ or $\mathbb{Q}$), as the matrix that has the cycles of $B$ as columns. 

\bigskip

Different classes of cycle bases can be considered. In \cite{Liebchen:2007} the authors characterize them in terms of their corresponding cycle matrices and present a \emph{Venn diagram} that shows their inclusion relations. Among these classes we can find the \emph{strictly fundamental} class. 

\bigskip

The \emph{length} of a cycle is its number of edges. The \emph{minimum cycle basis} (\emph{MCB}) problem is the problem of finding a cycle basis such that the sum of the lengths (or edge weights) of its cycles is minimum. This problem was formulated by Stepanec \cite{Stepanec:1964} and Zykov \cite{Zykov:1969} for general graphs and by Hubicka and Syslo \cite{Hubicka:1975} in the strictly fundamental class context. In more concrete terms this problem is equivalent to finding the cycle basis with the sparsest cycle matrix. In \cite{Kavitha:2009} a unified perspective of the problem is presented. The authors show that the \emph{MCB} problem is different in nature for each class. For example in \cite{Deo:1982} a remarkable reduction is constructed to prove that the \emph{MCB} problem is NP-hard for the strictly fundamental class, while in \cite{Horton:1987} a polynomial time algorithm is given to solve the problem for the undirected class. Some applications of the \emph{MCB} problem are described in \cite{Kavitha:2009,Horton:1987,Deo:1982, Berger:2004}.

\bigskip

A related problem not covered in the literature (as far as we know) is to consider the sparsity of the \emph{grammian matrix} of a cycle matrix. Let $B = (C_1,\dots,C_\nu)$ be a cycle basis with corresponding cycle matrix $\Gamma$. The grammian of $\Gamma$ is  $\hat{\Gamma} = \Gamma^t \Gamma$. We will denote $\hat{\Gamma}$ the \emph{cycle intersection matrix} of $B$. It is easy to check that the $ij$-entry of $\hat{\Gamma}$ is 0 if and only if the cycles $C_i$ and $C_j$ do not intersect (i.e.: they have no edges in common). It can be formulated as follows:

\medskip
\textbf{Problem}: let $G$ be a (directed) graph, find a cycle basis $B$ with corresponding cycle matrix $\Gamma$ such that the grammian $\hat{\Gamma} = \Gamma^t \Gamma$ is sparsest.

\bigskip

In this context the MSTCI problem corresponds to the particular case of bases that belong to the strictly fundamental class.

\subsection{An application}
Let $G = (V,  E)$ be a directed connected graph and $w:  E \rightarrow \mathbb{R}$ be an edge function. We call $w$ a \emph{discrete 1-form} on $G$. Integrating $w$ is the problem of finding a vertex function $x: V \rightarrow 
\mathbb{R}$ minimizing the error:

$$E(x) = \sum_{e_{ij} \in  E} ||dx(e_{ij}) - w(e_{ij})||^2.$$

where $dx:  E \rightarrow \mathbb{R}^n$ is defined as 

$$dx(e_{ij}) := x(v_j) - x(v_i).$$

on every directed edge $e_{ij} := v_i \rightarrow v_j$, and is called 
the \emph{differential} of $x$. Note that $w$ has the following property: $w(e_{ij}) = - w(e_{ji})$, where $e_{ji}$ is the same underlying edge $e_{ij}$ with opposite direction. 

\smallskip

Given some consistent enumeration of the vertices and edges, the integration 
problem can be expressed in a compact form:

$$\operatorname*{argmin}_x ||D\bf x-\bf w||_2^2.$$

where $D \in \{0,1,-1\}^{|E| \times |V|}$ is the \emph{directed 
incidence matrix} of $ G$, $\bf w \in \mathbb{R}^{|E|}$ is the evaluation of $w$ on the edges and $\bf x \in \mathbb{R}^{|V|}$ is the solution. From a geometric perspective $D\bf x$ can be visualized as the orthogonal projection of $\bf w$ onto the subspace generated by $D$. The rank of the directed incidence matrix is $|V|-1$, its kernel is generated by $\mathbbm{1} \in \mathbb{R}^{|V|}$, the vector of all 1's. This degree of freedom can be geometrically interpreted as a rigid translation in the solution space. If we fix the value of the 0-form at some vertex, we can eliminate this degree of freedom. More precisely, we can fix the value of the first component of our solution vector: $x(v_1) = 0$. This will be equivalent to eliminating the first column of $D$. Let $\hat{D}$ be this new matrix of dimension $|E| \times |V|-1$.

\smallskip

An alternative way to solve the integration problem is to extend $\hat{D}$ to a basis of $\mathbb{R}^{|E| \times |E|}$ and solve a linear system:

$$
M
\begin{bmatrix}
	\bf \hat x\\
	\bf y
\end{bmatrix} = 
\left[
\begin{array}{c|c}
\hat{D} & \hat{D}^{\perp} 
\end{array}\right]
\begin{bmatrix}
	\bf \hat x\\
	\bf y
\end{bmatrix}
=
\left[
\begin{array}{c|c}
\hat{D} & \Gamma
\end{array}\right]
\begin{bmatrix}
	\bf \hat x\\
	\bf y
\end{bmatrix}
=
\bf w.
$$

where $\bf \hat x$ is the solution vector $\bf x$ without its first component (because $x(v_1) = 0$) and $\hat{D}^{\perp}$ is a set of generators of the orthogonal complement of  $\hat{D}$. In this setting, a natural question is: how can we choose $\hat{D}^{\perp}$ such that $M$ is as sparse as possible? An answer is given by considering the \emph{cycle matrix} $\Gamma$ of the \emph{minimum cycle basis} $B$ of $G$. It is easy to check that the columns of $\Gamma$ are orthogonal to the columns of $\hat{D}$.

\smallskip

Yet another way of solving the integration problem is to consider the \emph{Gram} matrix (i.e.: $M^tM$) of $M$. More precisely, we can left multiply by $M^t$ in the previous equation:

$$
M^tM
\begin{bmatrix}
	\bf \hat x\\
	\bf y
\end{bmatrix} = 
\left[
\begin{array}{c|c}
\hat{D} & \Gamma
\end{array}\right]^t
\left[
\begin{array}{c|c}
\hat{D} & \Gamma
\end{array}\right]
\begin{bmatrix}
	\bf \hat x\\
	\bf y
\end{bmatrix}
=
\left[
\begin{array}{c|c}
\hat L & 0 \\
0 & \hat \Gamma
\end{array}
\right]
\begin{bmatrix}
	\bf \hat x\\
	\bf y
\end{bmatrix}
=
M^t\bf w.
$$

where $\hat L = \hat{D}^t \hat{D}$ is the lower right $|V|-1 \times |V|-1$ submatrix of the \emph{laplacian} matrix of $G$ and $\hat \Gamma = \Gamma^t \Gamma$ is the cycle intersection matrix of $B$. The same question can be formulated in this setting: how can we choose $B$ such that its corresponding cycle intersection matrix $\hat \Gamma$ is as sparse as possible? In the particular case where the cycle basis is in fact a strictly fundamental cycle basis, namely a cycle basis induced by a spanning tree, this is precisely the MSTCI problem.

\section{Preliminaries}
\subsection{Overview}

In the first part of this section we present some of the terms  
used in this article. Then, we define the notion of \emph{closest-point} 
and \emph{closest-point-set}. Finally, we show a convenient cycle 
partition.

\subsection{Notation}
Let $G=(V,E)$ be a graph and $T$ a spanning tree of $G$. We will then
refer to the edges $e\in T$ as \emph{tree-edges} and to the $e 
\in G-T$ ones as \emph{cycle-edges}.

\smallskip

Every cycle-edge $e$ induces a cycle in $T \cup \{e\}$, which we will 
call a \emph{tree-cycle}. We will name $C_T$ to the 
set of tree-cycles of $T$. 

\smallskip

The intersection of two tree-cycles is the set of edges 
of $T$ that belongs to both cycles. We will define three functions 
concerning the intersection of tree-cycles. 

\smallskip

The first is $\cap_{T}(\cdot, \cdot): C_T \times C_T \rightarrow \{0,1\}$ 

$$\cap_{T}(c_i, c_j) :=
	\begin{cases}
	  1 & c_i \cap c_j \neq \varnothing \land c_i \neq c_j. \\
	  0 & c_i \cap c_j = \varnothing \lor c_i = c_j. \\
	\end{cases}
$$

As every tree-cycle intersects with itself, the case $c_i = c_j$ is excluded to focus on non-trivial intersections. This consideration will simplify future computations.

\smallskip

The second is 
$\cap_{T}(\cdot): C_T \rightarrow \mathbb{N}$ 

$$\cap_{T}(c_i) := \sum_{c_j \in C_T} \cap_{T}(c_i, c_j).$$

We will call $\cap_{T}(c)$ the \emph{cycle intersection number} of 
$c$. Given a tree-cycle $c$ we will denote $\cap_{T,c}$ as the set of 
tree-cycles that have non-empty intersection with $c$. More precisely:

$$\cap_{T,c} \equiv \{c' \in C_T:  \cap_{T}(c, c') = 1 \}.$$

Note that $|\cap_{T,c}| = \cap_{T}(c)$.

\smallskip

In order to define the third function, consider $\mathscr{T}_G$ to be the 
set of spanning trees of $G$, therefore the definition will be as follows: 
$\cap_G: \mathscr{T}_G \rightarrow \mathbb{N}$

$$\cap_G(T) := \frac{1}{2}\sum_{c \in C_T} \cap_{T}(c).$$

We will call $\cap_G(T)$ the \emph{tree intersection number} of $T$. 

If the graph is clear from the context, 
we could remove the subindex and just write $\cap(T)$.

\bigskip

We shall call \emph{star} spanning tree to one that 
has one vertex that connects to all other vertices, and $K_n$ to the 
complete graph on $n$ nodes. If $G=(V,E)$ we will say that $|V| = n$ is 
the number of vertices of $G$, $|c| = k$ is the length of the cycle 
$c$ and $|p|$ is the length of the path $p$. Thus, $uTv$ will denote the 
unique path between $u,v \in V$ in the spanning tree $T$; and $d_T(v)$ will 
be the degree of $v \in V$ relative to it. Whereas $N(v)$ will be the set of neighbor nodes of $v \in V$ and finally, the terms ``node'' and ``vertex'' will be used interchangeably.

\subsection{Closest point}

In this section we prove the following simple fact: if $G=(V,E)$ is a 
connected graph, $T$ a spanning tree of $G$ and $c \in  C_T$ a 
tree-cycle, then for every node $v \in V$ there is a unique node $w 
\in c$ that minimizes the distance to $v$ in $T$. We shall denote that 
node $closest-point(v,c)$.

\bigskip

\begin{lemma}
Let $G=(V,E)$ be a connected graph, $T$ a spanning tree of $G$ and $c 
\in C_T$ a tree-cycle. Then for every node $v \in V$ 
there exists a unique node $w \in c$ such that 

$$|vTw| \leq |vTu| \ \forall u \in c.$$

Proof. The proof proceeds by contradiction. If $v \in c$, it is its own unique closest point. Suppose that $v \notin c$ and 
that there are two distinct nodes $w, w' \in c$ such that $|vTw| = 
|vTw'|  
\leq |vTu| \ \forall u \in c$, note that $w' \notin vTw$ and $w 
\notin vTw'$. We conclude that $vTw \cup wTw' \cup 
vTw'$ determines a cycle in $T$ which contradicts the fact that 
$T$ is a tree.$\square$
\end{lemma}

The uniqueness of the $closest-point(v,c)$ leads to the following 
definition.

\begin{definition} Let $G=(V,E)$ be a connected graph, $T$ a spanning tree of $G$ 
and $c \in C_T$ a tree-cycle, then the set of closest points to a node 
$w \in c$ is defined as follows:

$$closest-point-set(w,c) := \{v \in V-c: closest-point(v,c)=w \}.$$
\end{definition}

\subsection{Tree cycle intersection partition}
Now we define a partition of the set $\cap_{T,c}$. 
More precisely, let $G$ be a connected graph, $T$ a spanning tree of $G$ and $c \in 
C_T$ a tree-cycle. As mentioned above, the set $\cap_{T,c}$ is the set 
of tree-cycles that have non-empty intersection with $c$.

\bigskip

Let us consider any tree-cycle $c' \in \cap_{T,c}$ induced by a 
cycle-edge $e=(v,w)$. In this setting we can define the following 
partition:

\begin{itemize}
	\item \emph{Internal tree-cycles}: $c'$ is \emph{internal} if $v, w 
	\in c$.
	\item \emph{External tree-cycles}: $c'$ is \emph{external} if $v 
	\notin c$ and $w \in c$.
	\item \emph{Transit tree-cycles}: $c'$ is \emph{transit} if 
	$v, w \notin c$.
\end{itemize}

Let us denote the set of cycles of each type $\cap^i_{T,c}$, $\cap^e_{T,c}$, 
$\cap^t_{T,c}$, respectively. This partition will be convenient to 
simplify the computation of the \emph{intersection number} of $c$.

\subsection{Important remark}

There is an alternative point of view that may clarify some of the proofs of this article. Instead of considering tree-cycles as the central object, this point of view considers paths in the spanning tree. More precisely let $G$ be a connected graph, $T \in \mathscr{T}_G$ a spanning tree and $e_1,=(v_1, w_1), e_2 = (v_2,w_2) \in E$ two distinct cycle-edges with corresponding tree-cycles $c_1, c_2$. Then the following holds: $c_1 \cap c_2 = (v_1Tw_1) \cap (v_2Tw_2)$. Consequently, it is equivalent to consider tree-cycle intersections and intersection of paths in the spanning tree.

\section{Tree cycles of complete graphs}

\subsection{Overview}

In this section we analyze the complete graph case $G=K_n$ ($n \geq 3$). First we 
deduce a formula to compute the cycle intersection number. Then we 
prove that the tree-cycles of a star spanning tree achieve the minimum 
cycle intersection number. Finally, we conclude that the star spanning 
trees are the unique solutions of the MSTCI problem. 

\subsection{Cycle intersection number formula}

In this subsection we consider the problem of finding a formula to 
count tree-cycle intersections. More precisely, let $G=K_n$, 
$T$ a spanning tree of $G$ and $c$ a tree-cycle, we intend to derive a 
formula to calculate $\cap_{T}(c)$. 

\bigskip

The idea behind the formula is to consider the partition of 
$\cap_{T,c}$, defined in the previous section, and then by 
combinatorial arguments, compute the number of elements in each class.

\smallskip

We shall analyze in turn the three classes: $\cap^i_{T,c}$, 
$\cap^e_{T,c}$, $\cap^t_{T,c}$. In this section we will consider 
$c' \in \cap_{T}(c)$ to be a tree-cycle induced by a cycle-edge 
$e=(v,w)$.

\bigskip

The simplest case is the  
internal tree-cycles class: $\cap^i_{T,c}$. Let $c'$ be an
internal tree-cycle. By definition the nodes $v$ and $w$ belong 
to $c$, so the following holds: $(c' \cap T) \subset c$ because there is 
a unique path from $v$ to $w$ in $T$. So basically counting the number 
of internal tree-cycles reduces to counting the pairings on the nodes of 
$c$ excluding some obvious cases such as the pairing of a node with itself and with its neighbors in $c$. 
Then the number of internal tree-cycles is:

$$|\cap^i_{T,c}| = \frac{(k-3) k}{2}.$$
where $k$ is, as before, equal to $|c|$.
The quotient is obviously due to the fact that every cycle is counted 
twice.

\smallskip

Next we consider the class of external tree-cycles. Now let $c'$ 
be an external tree-cycle. In this case, by definition, either $v$ or $w$ belong to $c$. Without loss of generality (as we are 
considering undirected edges), suppose that $v \notin c$ and $w \in c$. 
Clearly $w \neq closest-point(v,c)$ because in that case $c' \cap c = 
\varnothing$ and consequently $c' \notin \cap_{T,c}$ which 
contradicts our hypothesis. Since the aforementioned is the only particular case that should be excluded, the number of external tree-cycles is:

$$|\cap^e_{T,c}| = (n-k) (k-1).$$
where $n=|V|$ is the number of vertices of $G$ and $k=|c|$ is the 
length of $c$.

Last we consider the class of transit tree-cycles. In this case the key 
observation depends on the $closest-point-set$ definition of the 
previous section. Let us define two classes of cycle-edges:

\begin{enumerate}
	\item A cycle-edge $e=(v,w)$ is called \emph{intraset cycle-edge} if 
	both $v,w \in closest-point-set(u_i,c)$ for some $u_i \in c$
	\item A cycle-edge $e=(v,w)$ is called \emph{interset cycle-edge} if 
	$v \in closest-point-set(u_i,c)$ and $w \in closest-point-set(u_j,c)$ 
	where $u_i, u_j \in c$ and $u_i \neq u_j$
\end{enumerate}

Then: 

\begin{itemize}
	\item Every \emph{intraset cycle-edge} induces a tree-cycle $c'$ such 
	that $c' \cap c = \varnothing$
	\item Every \emph{interset cycle-edge} induces a tree-cycle $c'$ such 
	that $c' \cap c \neq \varnothing$
\end{itemize}

So we should consider \emph{interset cycle-edges} or equivalently, the 
pairing of the nodes that are in different  
sets. Let $q_i = |closest-point-set(w_i,c)|$ be defined for all $w_i \in c$, then
the number of transit tree-cycles is:

$$|\cap^t_{T,c}| = \sum_{i<j} q_i q_j = \frac{1}{2} \sum_{i=1}^k q_i 
(n-k-q_i).$$

Finally, the intersection number formula is the aggregation of the three 
classes: 

$$\cap_{T}(c) = |\cap_{T,c}| = |\cap^i_{T,c}| + 
|\cap^e_{T,c}| + |\cap^t_{T,c}| = \frac{(k-3) k}{2} + (n-k) (k-1) + \frac{1}{2} \sum_{i=1}^k q_i 
(n-k-q_i).$$

where $n$ is the number of vertices of $G$, $k=|c|$ and 
$q_i = |closest-point-set(w_i,c)|$ for $w_i \in c$.

\subsection{Main result}

In this subsection we start by defining \emph{transitless} 
tree-cycles. Then we 
prove two lemmas. The first one shows that for every 
cycle $c \in G=K_n$ we can build a spanning tree $T$ such that $c$ is a 
tree-cycle of $T$ and the intersection number $\cap_{T}(c)$ is 
minimum. And the second one calculates the intersection number of 
tree-cycles of star spanning trees. Finally, we prove the main result of 
this section, namely that star spanning trees minimize $\cap(\cdot)$ in the 
case of complete graphs.

\bigskip

\begin{definition} Let $G=(V,E)$ be a connected graph, $T$ a spanning tree of 
$G$ and $c \in C_T$ a tree-cycle, we call $c$ a \emph{transitless} 
tree-cycle if $|\cap^t_{T,c}| = 0$.
\end{definition}

As an important remark, note that the number of elements in the 
internal and external classes of $c$ are independent of the spanning 
tree because they depend exclusively on $n=|V|$ and $k=|c|$. Thus, 
two spanning trees, $T_1$ and $T_2$, which have $c$ as a  
tree-cycle, induce an intersection number (for $c$) that only differs in 
the quantity of elements in their transit classes. We conclude that 
transitless tree-cycles have minimum intersection number.

\begin{lemma}\label{lemmatransitless}
Let $G=K_n$ and $c$ a cycle of $G$. Then, the following construction 
leads to a spanning tree $T$ that minimizes the intersection number of 
$c$: 

\begin{itemize}
	\item Let $e \in E$ such that $e \in c$.
	\item Let $v \in V$ such that $v \in c$.
	\item Define the set of edges of $T$ as follows: 
	$$E(T) = \{ e' \in E: e' \in (c - e)\} \cup \{ (v, w) \in E: w \in V \wedge w \notin c\}. $$
\end{itemize}

\smallskip

Proof. Note that $T$ is a spanning tree of $G$, and $c$ is a tree-cycle 
of $T$. Therefore, if we prove that 
$|\cap^t_{T,c}| = 0$ then the intersection number $\cap_{T}(c)$ is 
minimum.This is the case:

\begin{itemize}
	\item $|closest-point-set(w,c)| = 0$ $\forall w \in c, w \neq v$.
	\item $|closest-point-set(v,c)| = n-k$.
\end{itemize}

So $|\cap^t_{T,c}| = \sum_{i<j} q_i q_j = 0$.$\square$
\end{lemma}

\begin{lemma}
Let $G=K_n$ and let $T_s$ be a star spanning tree of $G$. Then the 
following property holds

$$\cap_{T_s}(c) = 2(n - 3).$$

for any tree-cycle $c$ of $T_s$.

\smallskip

Proof. Clearly the tree-cycles in $T_s$ have the same intersection 
number (by symmetry). Let $c$ be a tree-cycle of $T_s$. 
Note that $c$ is a triangle ($|c| = 3$), so the corresponding internal 
tree-cycle class is empty:  
$|\cap^i_{T,c}| = 0$. Being $c$ a transitless tree-cycle 
because its nodes are: the 
central node and two leaf nodes of $T_s$. Thus the external tree-cycle 
class is the only non-empty one: 

$$\cap_{T_s}(c) = |\cap^e_{T_s,c}| = 2(n - 3).$$

\end{lemma}

\begin{proposition}\label{propstar}
Let $G=K_n$ and let $T_s$ be a star spanning tree of $G$. Then the 
following property holds

$$\cap_{T_s}(\cdot) \leq \cap_{T}(\cdot).$$

where $T$ is any spanning tree of $G$.

\smallskip

Proof. We shall proceed by contradiction. Suppose that a 
spanning tree $T$ and a tree-cycle $c$ of $T$ exist such that:

$$\cap_{T}(c) < \cap_{T_s}(\cdot) = 2(n - 3).$$

We can assume that $c$ is transitless because, if this weren't the case, by 
Lemma \ref{lemmatransitless} we could build a spanning tree $T'$ such 
that $\cap_{T'}(c) < \cap_{T}(c)$. In this context the 
inequality can be expressed as 

$$\cap_{T}(c) = |\cap^i_{T,c}| + |\cap^e_{T,c}| = \frac{(k-3) k}{2} + (n-k) (k-1) < 2(n - 3).$$

Expanding and simplifying the expression we have

$$\frac{-1}{2} k^2 + (n-\frac{1}{2}) k - 3n + 6 < 0.$$

The roots of this quadratic polynomial are: $r_1 = 3$ and $r_2 = 
2(n-2)$.  
We should consider two cases depending on the relation of the roots:

\begin{enumerate}
	\item $r_1 < r_2$.
	\item $r_1 > r_2$.
\end{enumerate} 

The case $r_1 = r_2$ can be discarded because it leads to a 
fractional number of nodes ($n = \frac{7}{2}$).
In the first case the inequality holds for $k < r_1 = 3$ or $k > r_2 = 
2(n-2)$. The case $k < 3$ is an obvious contradiction since the size of 
the cycle must be $|c| = k \geq  
3$. The case $k > 2(n-2)$ combined with the fact that $k \leq n$ 
induces the following inequality

$$r_1 = 3 < r_2 = 2(n-2) < k \leq n.$$

which implies a contradiction: $3 < n < 4$, because $n$ is a positive 
integer.

The second case ($r_1 = 3 > r_2 = 2(n-2)$) implies $n < \frac{7}{2}$. So 
the only case that should be considered is $k = n = 3$ since $k \leq 
n$. However, this case makes the inequality false because $k = 3$ is a 
root of the quadratic polynomial.$\square$

\begin{corollary}\label{coromain}
Let $G=K_n$ and let $T_s$ be a star spanning tree of $G$. Then 
the following property holds

$$\cap(T_s) \leq \cap(T).$$
where $T$ is any spanning tree of $G$.

\smallskip

Proof. As expressed by Proposition \ref{propstar}, a tree-cycle of a 
star spanning tree has the minimum intersection number among all 
tree-cycles. Since any tree-cycle of a star spanning tree has the same 
intersection number, we conclude that the tree intersection number of 
a star spanning tree $\cap(T_s)$ is minimum among all spanning trees. 
$\square$
\end{corollary}

\smallskip

This corollary can be further improved to a strict inequality. In other 
words: star spanning trees are the unique minimizers of $\cap(T)$.

\smallskip

\begin{corollary}\label{coromain2}
Let $G=(V,E)=K_n$ where $|V| = n > 4$ and let $T_s$ be a star spanning 
tree of $G$. Then, the following property holds

$$\cap(T_s) < \cap(T).$$
where $T$ is any non-star spanning tree of $G$.

\smallskip

Proof. A careful reading of Proposition \ref{propstar} leads to the 
conclusion that the equality $\cap_{T_s}(c) = \cap_{T}(c)$ 
is achieved when $k$ is either $r_1=3$ or $r_2=2(n-2)$ 
(the roots of the quadratic polynomial). If $k = r_2 = 2(n-2)$, and 
taking into account that $3 \leq k \leq n$, we conclude that 
$\frac{7}{2} \leq n \leq 4$; this case is explicitly excluded from 
our hypotheses (in fact, it is not difficult to check that the three 
non-isomorphic spanning trees of $K_4$ have all the same tree 
intersection number).

\smallskip

The other possibility is $k = r_1 = 3$. As all the tree-cycles 
of $T_s$ fall into this category, it is enough to 
show that $T$ has a tree-cycle $c$ such that $|c| = k > 3$ to conclude 
our thesis. Let $w \in V$ be a node with maximum degree in $T$.
And let $d_T(w)$ denote the degree of $w$ in $T$ and 
$N(w)$ to the set of neighbors of $w$ in $T$.
Since $T$ is a non-star spanning tree then $2 \leq d_{T}(w) < n-1$. So there is a node $u \in V$ such that $u \notin N(w)$ in $T$, and there is a node 
$k \in N(w)$ such that $k \notin wTu$. Notice that the edge $e=(u,k) \notin T$ 
(as it would induce a cycle). Hence, it is a cycle-edge. Note that the 
tree-cycle induced by $e$ has length at least 4.$\square$
\end{corollary}

\end{proposition}

This result can be summarized in the following way: star spanning trees 
are the unique solutions for the MSTCI problem for complete graphs.

\section{Further generalization}

\subsection{Overview}

Now we explore some aspects of a slightly more general 
case, namely: the MSTCI problem in the context of a graph (not 
necessarily complete) $G=(V,E)$ that admits a star spanning tree 
$T_s$. In the first part we present a formula to calculate $\cap(T_s)$. 
In the second one we show that $\cap(T_s)$ is a local minimum in the 
domain of what we refer to as the ``spanning tree graph''. In the third 
we prove a result that suggests a general observation: the fact 
that a spanning tree of a graph $G$ is a solution for the MSTCI 
problem doesn't depend on an intrinsic property of $T$ but on the 
particular embedding of $T$ in $G$. Finally we conjecture a 
generalization of Corollary \ref{coromain}: $\cap(T_s) \leq \cap(T)$ 
for every spanning tree $T$ of $G$. 

\subsection{Formulas for star spanning trees}

In this subsection we present two formulas for graphs $G=(V,E)$ that admit 
a star spanning tree $T_s$. Let us denote $v \in 
V$ to the central node of $T_s$.

\bigskip

The first formula corresponds to the cycle intersection number of a 
tree-cycle  
$c = (u,v,w) \in C_{T_s}$, namely $\cap_{T_s}(c)$. Recall from the 
previous section that $c$ intersects neither transit nor internal 
tree-cycles: $|\cap^t_{T,c}| = 0$ and $|\cap^i_{T,c}| = 0$. So its 
non-empty intersections are the tree-cycles in the set  $\cap^e_{T,c}$. 
Note that the remaining  
incident edges to $u$ and $w$, are the only source of tree-cycles that 
have non-empty intersection with $c$. So the formula is 
straightforward:

$$\cap_{T_s}(c) = d(u) - 2 + d(w) - 2.$$
where $d(u)$ and $d(w)$ are the degrees of $u$ and $w$, resp.

Now we shall deduce a formula for the tree intersection number 
$\cap(T_s)$. Each edge $(u,v) \in T$ is contained in $d(u)-1$ tree-cycles, choosing two of them gives all the possible intersections equal to $(u,v)$. 
The formula is as follows:

$$\cap(T_s) = \sum_{u \in V - \{v\}} \binom{d(u) - 1}{2}
= \frac{1}{2} \sum_{u \in V - \{v\}} (d(u) - 1) (d(u) - 2) 
= \frac{1}{2} \sum_{u \in V - \{v\}} d(u)^2 - 3 d(u) + 2.$$

If we denote ${\bf d}$ as the degree vector of $G$, that is, a vector that 
has in the $i$-th component the degree of the $i$-th vertex. And taking 
into account that $\sum_{u \in V} d(u) = 2m$ then, the 
formula can be expressed as: 

$$\cap(T_s) = \frac{1}{2} [||{\bf d}||_{2}^2 - 6m - (n-1)(n-6)].$$

\subsection{Star spanning tree as a local minimum}

In this subsection we prove that a star spanning tree is a local minimum 
respect to the tree intersection number in the domain of the 
\emph{spanning tree graph}. We start by defining this second order 
graph of the original one $G=(V,E)$. Then we analyze the structure of 
the neighbors of a star spanning tree $T_s$. Finally 
we demonstrate the result by a bijection between tree-cycles to 
conclude that $\cap(T_s)$ is a local minimum.

\bigskip

\begin{definition} 
Let $G=(V,E)$ be a graph, and $S$ a subgraph of $G$. We denote 
as $e \leftrightarrow e'$ to the operation 
of replacing the edge $e \in S$ with the edge $e' \in G-S$.
We call this operation \emph{edge replacement on $S$}.
\end{definition}

\begin{definition} 
Let $G=(V,E)$ be a graph. We denote 
$SP_{_G}$ to the graph that has one node for every spanning 
tree of $G$  
and an edge between two nodes, if the corresponding spanning trees 
differ in exactly one edge replacement. We call this graph
the \emph{spanning tree graph of $G$}.
\end{definition}

\subsubsection{Neighborhood of $T_s$} Let $G=(V,E)$ be a graph that 
admits a star spanning tree $T_s$ with $v \in V$ as its center. Let 
$\alpha_{_{T_s}}$ be the node corresponding to $T_s$ in $SP_{_G}$ and 
let $\alpha_{_T}$ (with corresponding spanning tree $T$) be any 
neighbor of $\alpha_{_{T_s}}$. By definition $T_s$ and $T$ differ in 
exactly one edge replacement $e \leftrightarrow e'$ where $e=(v,w) \in 
T_s$ and $e'=(u,w) \in T$. Note that $T$ is exactly the same as $T_s$ 
except that the node $w$ is no longer connected to the central node $v$ 
but to the intermediate node $u$. This similar structure 
has direct consequences in the intersection numbers of both trees.

\bigskip

Now we prove the result of this section.

\begin{theorem}\label{teo:local}
Let $G=(V,E)$ be a graph that admits a star spanning tree $T_s$ with $v 
\in V$ as its center. Then, $T_s$ is a local minimum with respect to 
the tree intersection number in the domain of $SP_{_G}$. 

\smallskip

Proof. Let $T$ be a spanning tree corresponding to a neighbor of $T_s$ 
in $SP_{_G}$. We want to prove that $\cap(T_s) \leq \cap(T)$. 
Therefore, we shall proceed by defining a bijection between the tree-cycles of 
both trees $\{c \leftrightarrow d: c \in C_{_{T_s}} \wedge d 
\in C_{_T} \}$ such that $\cap_{_{T_s}}(c) \leq 
\cap_{_{T}}(d)$, this strategy clearly implies the thesis 
since by definition:

$$\cap(T_s) = \frac{1}{2}\sum_{c} \cap_{T_s}(c) \leq 
\frac{1}{2}\sum_{d} \cap_{T}(d) = \cap(T).$$

\smallskip

Let $e_{_{T_s}} \leftrightarrow e_{_T}$ with $e_{_{T_s}}=(v,w) \in T_s$ 
and $e_{_{T}}=(u,w) \in T$ be the edge replacement in 
$SP_{_G}$. Consider the following simple facts:

\begin{itemize}
	\item $e_{_{T}}$ is a cycle-edge in $T_s$, with corresponding 
	tree-cycle $c$.
	\item $e_{_{T_s}}$ is a cycle-edge in $T$, with corresponding 
	tree-cycle $d$.
	\item Except for $e_{_{T_s}}$ and $e_{_{T}}$, $T_s$ and $T$ have 
	the same set of cycle-edges. For every $e \in E - T_s - T$ we 
	denote $c_e$ and $d_e$ to the corresponding tree-cycles in $T_s$ 
	and $T$, resp.
\end{itemize}

According to this naming convention, we can define the following 
``natural'' bijection between tree-cycles:

$$\{c \leftrightarrow d\} \cup \{c_e \leftrightarrow d_e: e \in E - T_s 
- T\}.$$

In order to compare the intersection numbers of the bijected pairs it 
is convenient to distinguish the following partition:

\begin{itemize}
	\item Case 1: the pair induced by the edge replacement, $\{c 
	\leftrightarrow d\}$.
	\item Case 2: pairs induced by cycle-edges non-incident to $u$ nor 
	to $w$, $\{c_e \leftrightarrow d_e: e \in E - T_s - T \wedge u \notin 
	e \wedge w \notin e\}$.
	\item Case 3: pairs induced by cycle-edges incident to $u$ or 
	$w$, $\{c_e \leftrightarrow d_e: e \in E - T_s - T \wedge (u \in e 
	\lor w \in e) \}$.
\end{itemize}

Case 1 is the easiest: note that $c$ and $d$ are the same tree-cycle 
$(u,v,w)$, which is a transitless triangle, so its intersection number 
is determined by its external intersections: 

$$\cap_{_{T_s}}(c) = d(u) - 2 + d(w) - 2 = \cap_{_{T}}(d).$$

Case 2 is similar, let $e=(h,k)$ be a cycle-edge non-incident to $u$ 
or to $w$ and $c_e \leftrightarrow d_e$ its corresponding pair of 
bijected tree-cycles. Clearly $e$ determines the transitless triangle 
$(h,v,k)$ both in $T_s$ and $T$ and as $d_{_{T_s}}(h) = d_{_{T_s}}(k) = 
d_{_{T}}(h) = d_{_{T}}(k) = 1$, then every other edge incident to $h$ 
or $k$ induces a tree-cycle that intersects $(h,v,k)$. We conclude 
that: 

$$\cap_{_{T_s}}(c_e) = d(h) - 2 + d(k) - 2 = \cap_{_{T}}(d_e).$$

Case 3 is the one that should be analyzed more carefully. As we already 
know how to calculate intersection numbers of tree-cycles in $T_s$, 
we will focus on the tree-cycles of $T$ and divide this 
partition in two sub-partitions:

\begin{itemize}
	\item Case 3.1: pairs induced by cycle-edges incident to $u$, $\{c_e 
	\leftrightarrow d_e: e \in E - T_s - T \wedge u \in e \}$.
	\item Case 3.2: pairs induced by cycle-edges incident to $w$, $\{c_e 
	\leftrightarrow d_e: e \in E - T_s - T \wedge w \in e \}$.
\end{itemize}

In case 3.1 the situation is as follows: the cycle-edge $e=(u,k)$ 
defines the tree-cycle $c_e = d_e = (u,v,k)$ (both in $T$ and $T_s$). 
The important details are: 

\begin{itemize}
	\item $d_T(u) = 2$: $u$ induces $d(u) - 3$ intersections.
	\item $d_T(k) = 1$: $k$ induces $d(k) - 2$ intersections.
	\item $d_T(w) = 1$: $w$ induces $d(w) - 1$ intersections.
	\item $d(w) \geq 2$ since it is connected at least to $u$ and $v$ 
	in $G$.
	\item $w$ may have an incident cycle-edge connecting it to $k$, so 
	we should avoid counting twice that intersection.
\end{itemize}

Now we claim that 

$$\cap_{_{T}}(d_e) \geq d(u) - 3 + d(k) - 2 + d(w) - 1 - 
\epsilon(w,k) \geq d(u) - 2 + d(k) - 2 = \cap_{_{T_s}}(c_e).$$

where 

$$\epsilon(w,k)=
	\begin{cases}
	  1 & (w,k) \in E. \\
	  0 & otherwise. \\
	\end{cases}
$$

The inequality follows since $d(w) - 1 - \epsilon(w,k) \geq 1$.

\bigskip

In case 3.2 the situation is as follows: the cycle-edge $e=(w,h)$ 
defines the tree-cycle $d_e = (w,u,v,h)$ in $T$ and $c_e = (w,v,h)$ in
$T_s$. The important details are: 

\begin{itemize}
	\item $d_T(u) = 2$: $u$ induces $d(u) - 2$ intersections.
	\item $d_T(h) = 1$: $h$ induces $d(h) - 2$ intersections.
	\item $d_T(w) = 1$: $w$ induces $d(w) - 2$ intersections.
	\item $u$ may have an incident cycle-edge connecting it to $h$, so 
	we should avoid counting twice that intersection.
\end{itemize}

And we claim that 

$$\cap_{_{T}}(d_e) \geq d(w) - 2 + d(h) - 2 + d(u) - 2 - 
\epsilon(u,h) \geq d(w) - 2 + d(h) - 2 = \cap_{_{T_s}}(c_e).$$

The inequality follows since $d(u) - 2 - 
\epsilon(u,h) \geq 0$.

$\square$

\end{theorem}

\subsection{Intrinsic tree invariants}

In this subsection we consider the following question: is there any 
correlation between an \emph{intrinsic tree invariant} and the  
tree intersection number of the spanning trees for every graph? If so 
we could formulate an alternative characterization of the MSTCI 
problem expressed in terms of the invariant.

\bigskip

By \emph{intrinsic tree invariant} we denote a map $f: \mathscr{T} 
\rightarrow \mathbb{R}$ on the set of all trees. Of particular interest 
are the  degree-based topological indices \cite{Gutman:2013}. The 
topological index that motivated our question is the \emph{atom-bond 
connectivity} (ABC) index \cite{Estrada:1998}. As shown by 
\cite{Furtula:2009} the star trees are maximal among all trees respect 
to the ABC index. In the previous section we proved that in the 
complete graph the star spanning trees are minimal respect to the tree
intersection number. Consequently we can formulate a natural question: 
is there a negative correlation between the ABC index of the spanning 
trees and their corresponding intersection numbers?  

\bigskip

We will prove that the answer to our question is negative. Without loss 
of generality we will consider positive correlation (negative 
correlation is analogous). The underlying idea of the proof is as 
follows: suppose that there exists an intrinsic tree invariant $f: 
\mathscr{T} \rightarrow \mathbb{R}$ such that for every graph $G$ 
the intersection number $\cap(\cdot)$ is positively correlated with 
$f$. This can be expressed as: 

$$f(T_1) \leq f(T_2) \iff \cap_G(T_1) \leq \cap_G(T_2), \forall G, T_1, T_2.$$

According to this property if we consider two trees $T_1$ 
and $T_2$ and two graphs $G$ and $H$ such that $T_1, T_2 \in 
\mathscr{T}_{G}$ and $T_1, T_2 \in \mathscr{T}_{H}$, then this 
equivalence follows:

$$\cap_{G}(T_1) \leq \cap_{G}(T_2) \iff \cap_{H}(T_1) \leq \cap_{H}(T_2).$$

So it suffices to show that there exist $T_1$, $T_2$, $G$ and $H$ 
such that the equivalence is not satisfied to answer the question 
negatively.

\smallskip

First we prove a simple lemma regarding the tree intersection number of 
a spanning tree $T$ under the removal of a cycle-edge. Namely, if a 
cycle-edge $e$ is removed from $G$ then the tree intersection number of 
$T$ decreases exactly in the intersection number of its corresponding 
tree-cycle.

\begin{lemma}\label{lemma:edge_removal}
Let $G=(V,E)$ be a graph, $T \in \mathscr{T}_G$ a spanning tree, $e \in 
G-T$ a cycle-edge, and $c$ the corresponding tree-cycle, then the 
following holds: 

$$\cap_{G-e}(T) = \cap_{G}(T)-\cap_{T}(c).$$

\smallskip

Proof. As the spanning tree $T$ is the same in both $G$ and 
$G-e$, the remaining cycle-edges define the same tree-cycles so 
their pairwise intersection relations are identical. As $c$ is not a 
cycle in $G-e$ then the equality follows. $\square$

\end{lemma}

\begin{theorem}\label{intrinsic} There is no intrinsic tree invariant 
$f: \mathscr{T} \rightarrow \mathbb{R}$ positively 
correlated with the intersection number $\cap_G(\cdot)$ for every graph 
$G$.

\smallskip

Proof. We will proceed by contradiction: let $f$ be such an intrinsic 
tree invariant. Then by definition for arbitrary graphs $G$ and $H$ the 
following equivalences hold 

$$f(T_1) \leq f(T_2) \iff \cap_{G}(T_1) \leq \cap_{G}(T_2).$$

$$f(T_1) \leq f(T_2) \iff \cap_{H}(T_1) \leq \cap_{H}(T_2).$$

where $T_1, T_2 \in \mathscr{T}_{G}$ and $T_1, T_2 \in 
\mathscr{T}_{H}$. This in turn implies that

$$\cap_{G}(T_1) \leq \cap_{G}(T_2) \iff \cap_{H}(T_1) \leq \cap_{H}(T_2).$$

\smallskip

The proof will be based on showing two graphs and two spanning trees 
such that the latter equivalence is not valid. 

\begin{itemize}
	\item Let $G$ be the complete graph $K_n$.
	\item Let $H$ be the graph $K_n - \{e_{i,1}, \dots, e_{i,n-3}\}$.
	where the edges $e_{i,1}, \dots, e_{i,n-3}$ are $n-3$ edges incident 
	to some arbitrary node $v_i$. We will refer to $v_i$ as the 
	\emph{almost disconnected} node of $H$. Note that $d(v_i) = 2$.
	\item Let $T_1$ be the star spanning tree $T_s$.
	\item Let $T_2$ be the spanning tree defined as $T_s - \{e_i\} \cup 
	\{e_{i,j}\}$, where $e_i$ is the edge that connects some arbitrary 
	node $v_i$ (in $H$ this role will be played by the almost 
	disconnected node) to the center of the star and $e_{i,j}$ is an edge 
	that connects $v_i$ to a different node $v_j$.
\end{itemize}

It is easy to check that $T_1$ and $T_2$ are spanning trees of both $G$ and 
$H$. If we also suppose that $|V| = n > 4$ then by Corollary \ref{coromain2}

$$\cap_G(T_1) < \cap_G(T_2).$$

By the previous equivalence it is expected that $\cap_H(T_1) < 
\cap_H(T_2)$ as well. But we will show that this is not the case.

\smallskip

By a suitable labelling of the nodes of $H$ we can refer to: the center 
of the star spanning tree as $v_1$, the almost disconnected node of $H$ 
as $v_2$ and the other neighbor of $v_2$ as $v_3$. By Lemma 
\ref{lemma:edge_removal} we have that 

$$\cap_H(T_1) = \cap_{H-e_{_{2,3}}}(T_1) + \cap_{T_1}(c_{_{2,3}}).$$

$$\cap_H(T_2) = \cap_{H-e_{_{1,2}}}(T_2) + \cap_{T_2}(c_{_{1,2}}).$$

where $c_{_{2,3}}$ and $c_{_{1,2}}$ are the tree-cycles induced by 
$e_{_{2,3}}$ and $e_{_{1,2}}$ in $T_1$ and $T_2$, resp. The 
remaining tree-cycles corresponding to both trees are the same then

$$\cap_{H-e_{_{2,3}}}(T_1) = \cap_{H-e_{_{1,2}}}(T_2).$$

And this implies the following

$$\cap_H(T_1) - \cap_H(T_2) = \cap_{T_1}(c_{_{2,3}}) - 
\cap_{T_2}(c_{_{1,2}}).$$

It is an easy exercise to check that

$$\cap_{T_1}(c_{_{2,3}}) = \cap_{T_2}(c_{_{1,2}}) = d(v_3) - 2 = n - 3.$$

At this point we can conclude that 

$$\cap_H(T_1) = \cap_H(T_2).$$

Contradicting the fact that $f$ is positively correlated with the 
tree intersection number for every graph.$\square$
\end{theorem}

The underlying key fact of this result is that a spanning tree $T$ that 
solves the MSTCI problem for a graph $G$ does not depend on intrinsic 
properties of $T$ but on the embedding of $T$ in $G$.

\bigskip

Note that as an interesting side effect this demonstration shows that 
a star spanning tree is not necessarily a strict local minimum in the 
spanning tree graph (see previous subsection). 

\subsection{Intersection number conjecture}

In this subsection we present the conjecture 
$\cap(T_s) \leq \cap(T)$ for every spanning tree $T$ 
which generalizes Theorem \ref{teo:local}. Then we explore two ideas to 
simplify a hypothetical counterexample of the conjecture.
The first is based on the notion of \emph{interbranch} cycle-edge. We 
show that if a non-star spanning tree $T$ exists such that $\cap(T) < 
\cap(T_s)$, then the inequality must hold if we remove the interbranch 
cycle-edges. The second is based on the notion of \emph{principal 
subtree}. In this case we show that the inequality must hold for some 
principal subtree of $T$. These ideas 
will be of practical use in the next section.

\subsubsection{The conjecture statement}
We present below the conjecture that generalizes the case of complete 
graphs.

\bigskip

\begin{conjecture}\label{conj}
Let $G=(V,E)$ be a graph that admits a star spanning tree $T_s$, then 

$$\cap(T_s) \leq \cap(T).$$

for every spanning tree $T \in \mathscr{T}_G$.
\end{conjecture}

\bigskip

As an important remark, a demonstration of this result seems difficult 
if  
approached by a local-to-global strategy as in the complete graph 
case exposed previously.

\subsubsection{Counterexample simplification}

\bigskip

In this part we consider some ideas to simplify a hypothetical 
counterexample of Conjecture \ref{conj}. 

\bigskip

Below we define the notion of \emph{interbranch} cycle-edge.

\bigskip

\begin{definition} 
Let $G=(V,E)$ be a graph that admits a star spanning tree $T_s$ and let 
$v \in V$ be the center of $T_s$. Let $T \in 
\mathscr{T}_G$ be a spanning tree.
We call \emph{interbranch cycle-edge of $T$} to any 
cycle-edge of $T$, $e=(u,w)$, such that $closest-point(v,c) \neq u,w$,
where $c$ is the induced tree-cycle of $e$ in $T$.
\end{definition}

The intuition behind this definition is that the paths $vTu$ and $vTw$ 
belong to different branches with respect to $v$, more precisely, $vTu \not\subset vTw$ and $vTw \not\subset vTu$. The following lemmas show 
that if we can find a counterexample to the Conjecture \ref{conj} (i.e..: 
$\cap(T) < \cap(T_s)$) then we can build a simpler one 
removing the interbranch cycle-edges of $T$ from $G$.

\smallskip

\begin{lemma}\label{lemma:interbranch_edge}
Let $G=(V,E)$ be a graph that admits a star spanning tree $T_s$ with $v 
\in V$ as its center. Let $T \in \mathscr{T}_G$ be a spanning tree and $e = (u,w) \in \Delta_T$ an 
interbranch cycle-edge of $T$, then $e$ is a cycle-edge of $T_s$.

\smallskip

Proof. Since $v \neq u, w$ by definition of interbranch cycle-edge, then $u$ and $w$ are leaves of $T_s$ and consequently $e$ is a cycle-edge of $T_s$. $\square$

\end{lemma}

\begin{lemma}\label{lemma:interbranch_cycle}
Let $G=(V,E)$ be a graph that admits a star spanning tree $T_s$ with $v 
\in V$ as its center. Let $T \in \mathscr{T}_G$ be a spanning tree and $e = (u,w) \in \Delta_T$ an 
interbranch cycle-edge of $T$, then

$$\cap_{_{T_s}}(c) \leq \cap_{_{T}}(c').$$

where $c$ and $c'$ are the tree-cycles induced by $e$ in $T_s$ and $T$, resp.

\smallskip

Proof. By the intersection number formula we have that $\cap_{_{T_s}}(c) = d(u) - 2 + d(w) - 2$.

\smallskip

In order to prove that $\cap_{_{T_s}}(c) = d(u) - 2 + d(w) - 2  \leq \cap_{_{T}}(c')$ we have to consider the set of neighbors of $u$ and $w$ in $G$. We will consider only the set $N(w)$ because the same argument is valid for $N(u)$. Below we will show that $N(w)$ contributes with at least $d(w) - 2$ tree-cycles to $\cap_{_{T}}(c')$.

\smallskip

Let $h \not = u \in N(w)$ be the other neighbor of $w$ in $c'$(i.e. $(h,w) \in c'$). Note that the edge $(h,w) \in vTw$ because  the definition of interbranch cycle-edge requires that $closest-point(v,c') \not = w$. More concretely, $h$ is the immediate predecessor of $w$ in $vTw$. We intend to show that for every vertex $k \in N(w) - \{h,u\}$ there is a distinct tree-cycle in $T$ with non-empty intersection respect to c', thus achieving the claimed bound. We will consider the following cases:

\begin{enumerate}
	\item $vTw \subset vTk$.
	\item $vTk \subset vTw$.
	\item $vTw \not \subset vTk$ and $vTk \not \subset vTw$.
\end{enumerate}

In the first case $vTw$ is a subpath of $vTk$, then the edge $(v,k)$ is a cycle-edge of $T$ that determines a tree-cycle that contains the edge $(h,w)$.

\smallskip

In the second case $vTk$ is a subpath of $vTw$, then the edge $(w,k)$ is a cycle-edge of $T$ that determines a tree-cycle that contains the edge $(h,w)$.

\smallskip

In the third case there is no proper inclusion between $vTw$ and $vTk$, then the edge $(w,k)$ is a cycle-edge of $T$ that determines a tree-cycle that contains the edge $(h,w)$.

\bigskip

To check that the tree-cycles induced in this way by $N(w)$ and $N(u)$ are all distinct, note that in cases 2 and 3 the corresponding cycle-edges are incident either to $u$ or to $w$. And case 1 cannot occur simultaneously ($vTw \subset vTk$ and $vTu \subset vTk$) because $closest-point(v,c') \not = u, w$.

So the claimed inequality follows.$\square$

\end{lemma}

\begin{lemma}\label{lemma:reduction1}
Let $G=(V,E)$ be a graph that admits a star spanning tree $T_s$ with $v 
\in V$ as its center. Let $T \in \mathscr{T}_G$ be a spanning tree such 
that $\cap_{G}(T) < \cap_{G}(T_s)$ and let $\Delta_T$ be the set of 
interbranch cycle-edges of $T$, then

$$\cap_{_{G-\Delta_T}}(T) < \cap_{_{G-\Delta_T}}(T_s).$$

\smallskip

Proof. Let $e=(u,w) \in \Delta_T$, $c$ and $c'$ the tree-cycles induced by $e$ in $T_s$ and $T$, resp. By Lemma \ref{lemma:edge_removal} the following holds:

$$\cap_{G-e}(T_s) = \cap_{G}(T_s)-\cap_{T_s}(c).$$

$$\cap_{G-e}(T) = \cap_{G}(T)-\cap_{T}(c').$$

and by Lemma \ref{lemma:interbranch_cycle} we have that: 

$$\cap_{_{T_s}}(c) \leq \cap_{_{T}}(c').$$

we conclude that

$$\cap_{G-e}(T) = \cap_{G}(T) - \cap_{T}(c') < \cap_{G}(T_s) - \cap_{T_s}(c) = \cap_{G-e}(T_s).$$

Applying this edge removal for every edge in $\Delta_T$, the claimed inequality follows.$\square$

\end{lemma}

\begin{definition} 
Let $T=(V,E)$ be a rooted tree graph with root $v \in V$. Let $w \in 
N(v)$ then we call \emph{principal subtree with respect to w} to the 
subtree spanned by $v$ and the nodes $u \in V$ such that $w \in vTu$.
\end{definition}

The next lemma expresses the intersection number of a spanning tree 
(without interbranch cycle-edges) as the sum of the intersection number 
of its principal subtrees.

\begin{lemma}\label{lemma:partition}
Let $G=(V,E)$ be a graph that admits a star spanning tree $T_s$ with $v 
\in V$ as its center. Let $T$ be a spanning tree of $G$ without 
interbranch cycle-edges (i.e.: $\Delta_T = \varnothing$), then the 
following holds

$$\cap_G(T) = \sum_{w \in N(v)} \cap_{G_w}(T_w).$$

where $T_w$ is the principal subtree of $w \in N(v)$ 
considering $T$ as a rooted tree with $v$ as its root. And $G_w$ is the 
subgraph spanned by $T_w$.

\smallskip

Proof. As $\Delta_T=\varnothing$ there are no cycle-edges  
connecting any two such principal subtrees. This 
implies that the non-empty intersections between tree-cycles of $T$ must 
occur inside each subtree. This determines a partition of $C_T$ and the 
claimed  expression follows.$\square$
\end{lemma}

The following corollary, in line with Lemma \ref{lemma:reduction1}, 
further simplifies a hypothetical counterexample of Conjecture 
\ref{conj}.

\begin{corollary}\label{corollary:reduction2}
Let $G=(V,E)$ be a graph that admits a star spanning tree $T_s$ with $v 
\in V$ as its center. Let $T$ be a spanning tree of $G$ without 
interbranch cycle-edges (i.e.: $\Delta_T = \varnothing$) such that 
$\cap(T) < \cap(T_s)$ then

$$\cap(T_w) < \cap(G_w \wedge T_s).$$

for some $G_w$, where $T_w$ is the principal subtree of $w \in 
N(v)$ considering $T$ as a rooted tree with $v$ as its root; $G_w$ is 
the subgraph of $G$ spanned by $T_w$; $G_w \wedge T_s$ is the 
subtree of $T_s$ restricted to $G_w$, namely the intersection between 
$G_w$ and $T_s$.

\smallskip

Proof. First note that the $G_w$'s are edge disjoint since $\Delta_T = 
\varnothing$. This partition of the edges of $G$ also determines a 
partition of $T_s$ such that $\cap(T_s) = \sum_{w \in 
N(v)} \cap(G_w \wedge T_s)$. As the parts are in a natural bijective 
relation because they are the subtrees of $T$ and $T_s$ restricted to 
each $G_w$, we can express the intersection number of $T$ and $T_s$ as 
follows 

$$\cap(T) = \sum_{w \in N(v)} \cap(T_w) < \sum_{w \in N(v)} \cap(G_w 
\wedge T_s) = \cap(T_s).$$

And from the bijection we can deduce that 
$\cap(T_w) < \cap(G_w \wedge T_s)$ for some $G_w$.
$\square$
\end{corollary}

\section{Programmatic exploration}

\subsection{Overview}

In this section we present some experimental results to reinforce 
Conjecture \ref{conj}. We proceed by trying to find a counterexample based on our previous observations. In the first part, we focus on the complete analysis of small graphs, that is: graphs of at most 9 nodes. In the second part, we analyze larger families of graphs by random sampling instances.

\subsection{General remarks}

In the previous section we showed that the space of candidate 
counterexamples of Conjecture \ref{conj} can be reduced. The general 
picture is as follows: 

\begin{itemize}
	\item Let $G=(V,E)$ be a graph that admits a star spanning 
	tree $T_s$ with $v \in V$ as its center.
	\item In the case that we can find some non-star spanning tree $T$ of 
	$G$ such that $\cap(T) < \cap(T_s)$ then, we can ``simplify'' the instance by removing the interbranch cycle-edges with respect to $T$ in $G$ without affecting the inequality (see Lemma \ref{lemma:reduction1}).
	\item We can further reduce the instance by focusing on the case where 
	 $d_T(v) = 1$, that is: the degree of $v$ restricted to $T$ is 1 
	 (see Corollary \ref{corollary:reduction2}).
\end{itemize}

These considerations can be used to implement  algorithms to explore the 
space of spanning trees more efficiently, since the algorithms will 
generate instances in this `reduced' form instead of a brute force 
approach.

\subsection{Complete analysis of small graphs}

In this subsection we present an algorithm to explore the spanning tree 
 space. The algorithm proceeds by exhaustively analyzing all the 
 reduced graphs of a given number of nodes. The size of the space 
 increases exponentially with respect to the number of nodes, so it has 
 a major limitation: it can only be used to analyze small graphs. The 
 main part  is sketched in Algorithm \ref{alg:alg1}. 

\bigskip

The details of the algorithm are the following:

\begin{itemize}
	\item The input parameter $n$ is the number of nodes of the graphs to 
	explore.
	\item $GenerateAllTrees(n-1)$ is a function that returns 
	the list of all trees of $n-1$ nodes.
	\item $GenerateGraph(w,T')$ is a function that builds a graph $G$. 
	Based on the tree $T'$, it adds a new node ($v$), which will play the role 
	of the central node of a star spanning tree, and then the edge 
	$(v,w)$, to define our candidate tree counterexample $T$. Finally adds all the other edges that link $v$ to the rest of the nodes to obtain $G$. It 
	returns the graph $G$ and ($\bar{\Delta}$) the set of ``possible'' 
	non-interbranch cycle edges.
	\item $IntersectionNumber(\phi, G)$ is a function that calculates 
	the intersection number of $T$ in $G \cup \phi$, where $\phi 
	\subset \bar{\Delta}$ is a subset of supplementary edges of $G$.
	\item $StarIntersectionFormula(\phi, G)$ is a function that 
	calculates the intersection number of the star spanning tree in $G 
	\cup \phi$.
	\item The algorithm finds a counterexample of the conjecture if \\
	$IntersectionNumber(\phi, G) < StarIntersectionFormula(\phi, G)$. 
\end{itemize}

Note that the analyzed graphs are reduced in the sense previously explained. The cycle-edges are non-interbranch by construction and 
$d_T(v) = 1$ since $v$ is only connected to $w$ in $T$ (i.e.. there is a 
single principal subtree). As the 
algorithm iterates over all possible spanning subtrees $T'$ and all the 
combinations of possible non-interbranch cycle-edges, every instance is 
guaranteed to be explored at least once.

\begin{algorithm}
    \caption{CounterexampleSearch($n$)}
	\begin{algorithmic}
		\State $\mathscr{T} \gets GenerateAllTrees(n-1)$
		\ForEach {tree $T' \in \mathscr{T}$}
			\ForEach {node $w \in T'$}
				\State $G, \bar{\Delta} \gets GenerateGraph(w,T')$
				\ForEach {subset $\phi \subset \bar{\Delta}$}
					\State \textbf{check} $(IntersectionNumber(\phi, G) <$ 
					\State $StarIntersectionFormula(\phi, G))$
				\EndFor
			\EndFor
		\EndFor
	\end{algorithmic}
	\label{alg:alg1}
\end{algorithm}

\begin{table}
	\caption{Results for Small Instances}
	\centering
	\begin{tabular}{cr}
		\toprule
		Nodes & Instances (approx.) \\
		\midrule
		4 & 5 \\
		5 & 33 \\
		6 & 251 \\
		7 & 4200 \\
		8 & 125000 \\
		9 & 7900000 \\
		\bottomrule
		\label{tab:small-inst}
	\end{tabular}
\end{table}

\bigskip

In order to generate all non-isomorphic trees of $|V|-1$ nodes, we used the 
package \emph{nauty} \cite{McKay:2014}.

\bigskip

The proposed algorithm did not find a counterexample of the 
intersection conjecture. Table \ref{tab:small-inst} shows the size of 
the experiments. Column \emph{Nodes} is the number of nodes of the 
graph family, i.e.: $|V|$. Column \emph{Instances} is the number of 
instances processed.

\subsection{Random sampling of large graphs}

In this section we present another algorithm to explore the 
spanning tree space. The strategy in this case is to sample reduced 
graphs of a given number of nodes. The main part is sketched in 
Algorithm \ref{alg:alg2}.

\bigskip

The details of the algorithm are the following:

\begin{itemize}
	\item The input parameters are: $n$ the number of nodes of the graphs 
	and $k$ the size of the sample.
	\item $GenerateRandomTree(n)$ is a function that returns 
	a random tree $T$ of $n$ nodes, where the node $v$, that will play the 
	role of center of the star, has degree 1 restricted to $T$.
	\item $GenerateGraph(T)$ is a function that builds a reduced graph 
	$G$. Based on the tree $T$, it adds all the edges that link $v$ to 
	the rest of the nodes to obtain $G$. It returns the graph $G$ and 
	($\bar{\Delta}$) a random set of non-interbranch cycle edges.
	\item $IntersectionNumber(\phi, G)$ same as algorithm \ref{alg:alg1}. 
	\item $StarIntersectionFormula(\phi, G)$ same as algorithm 
	\ref{alg:alg1}.
	\item The algorithm finds a counterexample of the conjecture if: \\
	$IntersectionNumber(\phi, G) < StarIntersectionFormula(\phi, G)$. 
\end{itemize}

\begin{algorithm}
    \caption{CounterexampleRandomSearch($n, k$)}
	\begin{algorithmic}
		\For{\texttt{i := 1..k}}
			\State $T \gets GenerateRandomTree(n)$
			\State $G, \bar{\Delta} \gets GenerateRandomGraph(T)$
			\State \textbf{check} $(IntersectionNumber(\phi, G) <$ 
			\State $StarIntersectionFormula(\phi, G))$
		\EndFor
	\end{algorithmic}
	\label{alg:alg2}
\end{algorithm}

\begin{table}
	\caption{Results for Random Instances}
	\centering
	\begin{tabular}{cr}
		\toprule
		Nodes & Instances \\
		\midrule
		25 & 3000000 \\
		50 & 300000 \\
		100 & 30000 \\
		200 & 15000 \\
		400 & 300 \\
		\bottomrule
		\label{tab:random-inst}
	\end{tabular}
\end{table}

\bigskip

We used a uniformly distributed random number generator. To generate 
trees we used a simple algorithm that randomly connects a  
new node to an already connected tree. The non-interbranch cycle-edge 
set is built by associating a \emph{Bernoulli} trial to each  possible 
edge. To achieve some diversity for each tree we built three different sets to obtain sparse, medium and dense ones based on corresponding probabilities $0.1$, $0.5$, $0.9$.

\bigskip

The proposed algorithm did not find a counterexample of the 
intersection conjecture. Table \ref{tab:random-inst} shows the size of the experiments. Column \emph{Nodes} is the number of nodes of the graph family, i.e.: $|V|$. Column \emph{Instances} is the number of instances processed.

\section{Conclusion} 

In this article we introduced the \emph{Minimum Spanning Tree Cycle 
Intersection} (MSTCI) problem.

\bigskip

We proved by enumerative 
arguments that the star spanning trees are the unique solutions of the 
problem in the context of complete graphs.

\bigskip

We conjectured a generalization to the case of graphs (not necessarily 
complete) which admit a star  spanning tree. In this sense we showed that such tree is a 
local minimum in the domain of the \emph{spanning tree graph}. We 
deduced a closed  formula for the tree intersection number of star spanning 
trees in this setting. We proposed two  
ideas to reduce the search space of a counterexample of the conjecture. Those ideas were the basis of two strategies to programmatically explore the 
space of solutions in the pursuit of a counterexample. The negative 
result of the experiments suggests that the  
conjecture is well posed. Unlike the complete graph context, in this 
slightly more general case star spanning trees are not unique; there 
are other spanning trees $T$ such that $\cap(T_s) = \cap(T)$.

\bigskip

We proved a general result that shows that 
spanning trees that solve the MSTCI problem don't depend on some 
intrinsic property but on their particular embedding in the ambient 
graph.

\bigskip

An interesting direction of research is to consider the MSTCI
problem for other families of graphs, i.e..: graphs that do not admit a 
star spanning tree. Of particular interest for us is the class of triangular 
meshes, i.e..: graphs that model the immersion of compact surfaces in 
the 3D Euclidean space.

\bigskip

Another interesting line of research is to analyze the   complexity class of the MSTCI problem. In case of 
belonging to the NP-hard class, it will  be necessary to find 
approximate, probabilistic and heuristic algorithms.

\bigskip

In the introduction of this article we mentioned that the MSTCI problem is a particular case of finding a cycle basis with sparsest cycle intersection matrix. Another possible analysis would be to consider this in the context of the cycle basis classes described in \cite{Liebchen:2007}.

\section{Acknowledgments} 

The authors are grateful to the anonymous reviewers for their careful reading, constructive corrections and valuable comments on this paper, which have considerably improved its presentation.

\bigskip

\bibliographystyle{elsarticle-num-names}
\bibliography{mstci}

\end{document}